\documentclass[11pt]{article}
\usepackage{amsmath,amssymb,color,epsfig,cite}
\usepackage{graphicx}
\usepackage{subfigure}
\usepackage{setspace}

\usepackage{amsfonts}

\newcommand{\hoch}[1]{$\, ^{#1}$}

\makeatletter
\@addtoreset{equation}{section}
\makeatother

\newcommand{\be}{\begin{equation}}
\newcommand{\ee}{\end{equation}}
\newcommand{\bea}{\setlength\arraycolsep{2pt} \begin{eqnarray}}
\newcommand{\eea}{\end{eqnarray}}
\newcommand{\nn}{\nonumber}

\def\ft#1#2{{\textstyle{\frac{\scriptstyle #1}{\scriptstyle #2} } }}
\def\fft#1#2{{\frac{#1}{#2}}}

\def\0{{\sst{(0)}}}
\def\1{{\sst{(1)}}}
\def\2{{\sst{(2)}}}
\def\3{{\sst{(3)}}}
\def\4{{\sst{(4)}}}
\def\5{{\sst{(5)}}}
\def\6{{\sst{(6)}}}
\def\7{{\sst{(7)}}}
\def\8{{\sst{(8)}}}
\def\sst#1{{\scriptscriptstyle #1}}

\def\del{{\partial}}

\def\Ei{{\hbox{Ei}}}

\begin{document}

\begin{flushright}
\hfill{MI-TH-1744}

\end{flushright}

\begin{center}
{\large {\bf DC Conductivities with Momentum Dissipation in Horndeski Theories}}

\vspace{10pt}
Wei-Jian Jiang\hoch{1}, Hai-Shan Liu\hoch{2,3}, H. L\"u\hoch{4} and C.N. Pope\hoch{3,5}

\vspace{10pt}

\hoch{1}{ \it Zhejiang Institute of Modern Physics, Zhejiang University, Hangzhou 310058, China  } \\

\hoch{2}{\it Institute for Advanced Physics \& Mathematics,\\
Zhejiang University of Technology, Hangzhou 310023, China}

\vspace{10pt}

\hoch{3}{\it George P. \& Cynthia Woods Mitchell  Institute
for Fundamental Physics and Astronomy,\\
Texas A\&M University, College Station, TX 77843, USA}

\vspace{10pt}

\hoch{4}{\it Department of Physics, Beijing Normal University,
Beijing 100875, China}

\vspace{10pt}

\hoch{5}{\it DAMTP, Centre for Mathematical Sciences,
 Cambridge University,\\  Wilberforce Road, Cambridge CB3 OWA, UK}

\vspace{30pt}

\underline{ABSTRACT}

\end{center}

In this paper, we consider two four-dimensional Horndeski-type
gravity theories with scalar fields that give rise to solutions with
momentum dissipation in the dual boundary theories.
Firstly, we study Einstein-Maxwell theory with a Horndeski axion term and
two additional free axions which are responsible for momentum dissipation.
We construct static electrically charged  AdS planar black hole solutions in
this theory and calculate analytically the holographic DC conductivity of
the dual field theory. We then generalize the results to include magnetic
charge in the black hole solution. Secondly, we analyze Einstein-Maxwell
theory with two Horndeski axions which are used for momentum dissipation.
We obtain AdS planar black hole solutions in the theory and we calculate
the holographic DC conductivity of the dual field theory. The theory has
a critical point $\alpha + \gamma \Lambda = 0$, beyond which
the kinetic terms of the Horndeski axions become ghost-like.
The conductivity as a function of temperature behaves qualitatively
like that of a
conductor below the critical point, becoming semiconductor-like
at the critical point.  Beyond the critical point, the ghost-like nature of
the Horndeski fields is associated with the onset of unphysical
singular or negative conductivities.  Some further generalisations of the
above theories are considered also.

\vfill {\footnotesize Emails: hsliu.zju@gmail.com \ \ \ mrhonglu@gmail.com\ \ \
pope@physics.tamu.edu}

\thispagestyle{empty}

\pagebreak

\tableofcontents
\addtocontents{toc}{\protect\setcounter{tocdepth}{2}}



\section{Introduction}

Gauge/Gravity duality has served as a powerful tool in understanding the
phenomena of strongly coupled systems in condensed matter physics
\cite{ggd1,ggd2,ggd3}. Especially, much attention has been paid to
the holographic description of systems with momentum relaxation.
Such systems with broken translational
symmetry are needed in order to give a realistic description of
materials in many condensed matter systems.

Since momentum is conserved in a system with translational symmetry,
a constant electric field can generate a charge current without current
dissipation in the presence of non-zero charge density. Thus, the
conductivity of the system would become divergent at zero frequency.
In more realistic condensed matter materials, the momentum is not
conserved due to impurities or a lattice structure, thus leading to
a finite DC conductivity.

In the context of holography, there are various ways to achieve momentum
dissipation, such as periodic potentials, lattices and breaking
diffeomorphism invariance
\cite{lat1,lat2,lat3,lat4,lat5,lat6,lat7,lat8,EMDC,gels1,gels2,sels,mg1,mg2,mg3}. Among these,
the model in \cite{EMDC} is particular simple.  It comprises
an Einstein-Maxwell theory together with a set of minimally-coupled
massless scalar fields that have linear dependence on the boundary
coordinates. These axionic scalars preserve the homogeneity of the bulk
stress tensor, since they have no mass terms or interactions that would
break translational invariance.

In this paper, we shall generalise the models with momentum dissipation
that were constructed in \cite{EMDC} by introducing
non-minimal Horndeski type couplings of some of the scalar fields to gravity.
The Horndeski theories were first constructed in the 1970s \cite{hd1}, and
they have received much attention recently through their
application to cosmology in Galileon theories (see, for example,
\cite{hd2}).  A
characteristic feature of Horndeski theories is that although terms in their
Lagrangians involve more than two derivatives,  the field equations and
the energy-momentum tensor involve no higher than second derivatives of the
fields.   This is analogous to the situation in Lovelock gravities \cite{ll}.

Specifically, we shall generalise the model in \cite{EMDC} in
two parallel ways. Firstly, in section 2, we shall consider a
Horndeski extension
of an Einstein-Maxwell plus scalar theory in which two minimally-coupled
axions that provide the momentum dissipation are supplemented
by a third axion with a non-minimal Horndeski coupling.  Although this axion
has a significant effect in terms of modifying the geometrical structure
of the black hole
background, we find that the DC conductivity in the boundary theory is
essentially unaltered, at least if one expresses the result as a function
of the black hole horizon radius.
In section 3, we
shall consider instead an Einstein-Maxwell theory with Horndeski couplings to
the two axions that provide the momentum dissipation.  Here, we find that
the effects of the non-minimal Horndeski couplings are much more
substantial, and in fact as the strength of the non-minimal term is increased
to a critical value, the qualitative behaviour of the conductivities as a
function of temperature changes.  Below the critical coupling the
high-temperature behaviour is similar to that of a metal, whilst at the
critical coupling the behaviour becomes more like that of a
semiconductor.
 We summarize our results in section 4. In appendix, we extend the theories
and solutions that we studied in the main text to arbitrary spacetime
dimensions.

\section{Momentum dissipation with Horndeski term}

\subsection{Electrical black hole}

In this section, we consider AdS planar black holes of Horndeski theory in four dimensions.
The solutions have been constructed in \cite{hs1,hs2}, and the thermodynamics have been studied in \cite{th1,th2}.  In these solutions, the Horndeski axion $\chi$ depends on the radial coordinate.  In order to achieve momentum dissipation, we include two additional free axions $\phi_i$ as in \cite{EMDC}:
\bea
&&I = \fft{1}{16\pi}\int d^4 x\sqrt{-g}\, L\,, \cr
&&L = \kappa\Big(R-2\Lambda - \ft14 F^2 -\ft12 \sum_{i=1}^2
(\del\phi_i)^2 \Big)-
\ft{1}{2}(\alpha g^{\mu\nu}-\gamma G^{\mu\nu})
\, \del_\mu\chi\, \del_\nu\chi\,,
\eea
where $\kappa$, $\alpha$, $\gamma$ are coupling constants,
$G_{\mu\nu} \equiv R_{\mu\nu} - \fft12 R g_{\mu\nu}$ is the Einstein tensor,
and $F=dA$ is the electromagnetic field strength. The equations of
motion with respect to the metric $g^{\mu\nu}$, the Maxwell
potential $A_\mu$, the Horndeski scalar $\chi$ and the axions
$\phi_i$ are given by
\bea
 && \kappa (G_{\mu\nu} +\Lambda g_{\mu\nu} - \ft12 F_{\mu\nu}^2 + \ft18 F^2 g_{\mu\nu}) \cr
  && - \fft \kappa 2 (\partial_\mu \phi_1 \partial_\nu \phi_1 + \ft 12 \partial_\mu \phi_2 \partial_\nu \phi_2) + \fft \kappa 4 \big( (\partial \phi_1)^2 + (\partial \phi_2)^2  \big) g_{\mu\nu} \cr
&& -\ft12\alpha \Big(\partial_\mu \chi \partial_\nu \chi - \ft12 g_{\mu\nu} (\partial\chi)^2\Big)-\ft12\gamma \Big(\ft12\partial_\mu\chi \partial_\nu \chi R - 2\partial_\rho
\chi\, \partial_{(\mu}\chi\, R_{\nu)}{}^\rho \cr
&&- \partial_\rho\chi\partial_\sigma\chi\, R_{\mu}{}^\rho{}_\nu{}^\sigma -
(\nabla_\mu\nabla^\rho\chi)(\nabla_\nu\nabla_\rho\chi)+(\nabla_\mu\nabla_\nu\chi)
\Box\chi + \ft12 G_{\mu\nu} (\partial\chi)^2\cr
&&-g_{\mu\nu}\big[-\ft12(\nabla^\rho\nabla^\sigma\chi)
(\nabla_\rho\nabla_\sigma\chi) + \ft12(\Box\chi)^2 -
  \partial_\rho\chi\partial_\sigma\chi\,R^{\rho\sigma}\big]\Big) =0\,,\nn\\
&&\nabla_\mu \big( (\alpha g^{\mu\nu} - \gamma G^{\mu\nu})
\nabla_\nu\chi\big)=0\,,\qquad  \nabla_\nu F^{\nu\mu}  =  0\,,
\qquad \square\phi_i   =  0 \,. \label{Heoms}
\eea
One of the remarkable properties of a Horndeski theory is that each field
has no higher than second-derivative terms in the equations of motion,
even though the Lagrangian involves larger numbers of derivatives  (up to
four derivatives, in our case).  Although terms quadratic in
second-derivatives are present, linearised
perturbations around a background will involve at most second-order linear
differential equations,  and thus can be ghost free.

We are interested in static planar black hole solutions in this paper.
In this section, we shall take the Horndeski axion $\chi$ to depend only
on the radial coordinate,  whilst the two additional
axions $\phi_i$ span the planar directions:
\bea
&&ds^2 = - h(r) dt^2 + \fft{dr^2}{f(r)} + r^2 dx^i dx^i \,, \nn\\
&&\chi = \chi(r) \,, \quad A=a(r) \,dt\,, \qquad \phi_1 = \lambda x_1
\,, \qquad \phi_2 = \lambda x_2 \,, \label{ansatz}
\eea
where $\lambda$ is a constant.
The Maxwell equation can be used to express the electrostatic potential
in terms of the metric functions, as
\be
a' = \fft{q}{r^2} \sqrt{\fft h f}  \,,
\ee
where $q$ is an integration constant, parameterising the electric charge,
and a prime denotes a derivative with respect to $r$. The equation of motion
for the Horndeski scalar $\chi$ can
then be written as
\be
\Big( \sqrt{\fft{f}{h}}
\Big( \gamma \big( r f h' + f h\big)
    -\alpha r^2 h \Big)\chi'\Big)'=0\,.
\ee
Following \cite{hs1,hs2}, we focus on the special class of solutions
obtained by taking
\be
\gamma f \big( r  h' +  h\big)
   -\alpha r^2 h=0\,.
\ee
With this, we can solve the Einstein equations and obtain the
black hole solution
\bea
a &=&a_0-\fft{q}{r} + \fft{\kappa q^3}{30 g^2(4\kappa + \beta \gamma)\,r^5} + \fft{\kappa q \lambda^2}{9(4 \kappa +  \beta \gamma ) g^2 r^3}\,,\cr
\chi' &=& \sqrt{\beta - \fft{\kappa (q^2 + 2 \lambda^2 r^2)}{6\gamma g^2 r^4}}\,\fft{1}{\sqrt{f}}\,,\qquad
f= \fft{36(4\kappa + \beta\gamma)^2 g^4r^8}{\big(\kappa (q^2 + 2 \lambda^2 r^2) -6 (4\kappa + \beta\gamma) g^2 r^4   \big)^2 }\, h\,,\cr
h&=& g^2 r^2 - \fft{\mu}{r} + \fft{\kappa q^2}{(4\kappa + \beta\gamma)r^2} -
\fft{\kappa^2 q^4}{60(4\kappa + \beta\gamma)^2g^2 r^6} \cr
&& \qquad  \qquad - \fft{2 \kappa \lambda^2}{4 \kappa +  \beta \gamma} - \fft{\kappa^2 \lambda^4}{3 g^2r^2  (4 \kappa + \beta \gamma )^2} - \fft{\kappa^2 q^2 \lambda^2}{9 (4 \kappa + \beta \gamma)^2 g^2 r^4} \,, \label{esl}
\eea
where the parameters are such that
\be
\alpha= 3 g^2 \gamma\,,\qquad
\Lambda = - 3 g^2 \big(1 + \fft{\beta\gamma}{2\kappa}\big)\,.
\ee
The solution has non-trivial integration constants $\mu$, $q$ and $\lambda$,
together with a pure gauge parameter $a_0$.  The Hawking temperature can be
calculated by standard methods, and is given by
\be
T = \frac{ 6 g^2 r_0^4 (\beta  \gamma +4 \kappa ) -   \kappa ( q^2 + 2  \lambda ^2 r_0^2 )}{8 \pi  r_0^3 (\beta  \gamma +4 \kappa )} \,,
\ee
where $r_0$ is the radius of event horizon, which is the largest root of
$h(r) = 0$.

\subsection{DC conductivity}

There are many ways to compute the holographic conductivities.
For the DC conductivity, a simple method makes us of the ``membrane
paradigm'' \cite{mg2, DChong, DC1, DC2, DC3, DC4, DC5}.  The key
point is to construct a radially conserved current, which allows one
to read off the holographic boundary properties in terms of the
black hole horizon data.  Here, we shall follow the
procedure described in \cite{DC2}.

  We consider perturbations around the black hole solutions, of the form
\be
\delta g_{tx_1} = r^2 \psi_{tx} \,, \quad \delta g_{rx_1} = r^2 \psi_{rx} 
\,, \quad
\delta A_{x_1} = - E t + a_x \,, \quad \delta \phi_1 = \fft{\Phi}{\lambda} \,.
\ee
The equation of motion for the vector field $
\partial_r (\sqrt g F^{rx}) = 0$
implies that we can define a radially-conserved current
\be
J = \kappa \sqrt g  F^{rx_1}.
\ee
Explicitly, this current is given by
\be
J = \frac{\kappa \Big( f a_x' \left(-6 g^2 r^4 (\beta  \gamma +4 \kappa )+\kappa  q^2+2 \kappa \lambda ^2 r^2\right)-6 g^2 q r^4 \psi_{tx}
(\beta  \gamma +4 \kappa )\Big) }{6 g^2 r^4 (\beta  \gamma +4 \kappa )} \,,
\ee
and it obeys $\del J/\del r=0$.

The Einstein equations imply\footnote{Note that the perturbation
$\psi_{rx}$ is non-dynamical, and could in fact be removed by a
coordinate transformation. We choose to keep it here in order to make the
presentation parallel with the one we shall give below when a magnetic field
is turned on, since in that case one cannot remove the analogous
perturbations by means of a coordinate transformation.}
\bea
&& \frac{f \left(\Phi ' -  \lambda ^2 \psi_{rx} \right) \left(6 g^2 r^4 (\beta  \gamma +4 \kappa )-\kappa ( q^2 + 2 \lambda ^2 r^2) \right)}{6 g^2  r^2 (\beta  \gamma +4 \kappa )} = E  q \,, \cr
&&f  \left(4 \kappa  q a_x'+r^3
(\beta  \gamma +4 \kappa ) \left(r \psi_{tx}''+4 \psi_{tx}'\right)\right)= \fft{24 \kappa g^2 \lambda ^2 r^6 (\beta  \gamma +4 \kappa ) \psi_{tx}}{\left(6 g^2 r^4 (\beta  \gamma +4 \kappa )-
\kappa ( q^2+2 \lambda ^2 r^2)\right)}    \,.\label{psi}
\eea
Regularity on the horizon requires that
\be
a_x' = - \fft{E}{\sqrt {hf}} + {\cal O} (1) \,.
\ee
The last equation in (\ref{psi}) shows that near horizon,
\be
\psi_{tx} = -\frac{E  q}{\lambda ^2 r_0^2}  + {\cal O} (r-r_0) \,.
\ee
With these, we can evaluate the current on the horizon, finding
\be
J = \kappa ( 1 + \frac{q^2}{\lambda ^2 r_0^2}) E \,,
\ee
and hence the conductivity is given by
\be
\sigma = \fft{\partial J}{\partial E} = \kappa ( 1 + \frac{ q^2}{\lambda ^2 r_0^2}) \,. \label{cond}
\ee

Interestingly, even though the theory we are considering here, and its
black hole solutions, are much more complicated than
the Einstein-Maxwell theory with linear axions that was studied in
\cite{EMDC},
the Horndeski scalar $\chi$  does not explicitly contribute to the
conductivity when $\sigma$ is expressed in terms of $r_0$,  and hence
the result (\ref{cond}) is the same as in \cite{EMDC}. Of course,
the Horndeski term modifies the relation between the temperature and
$r_0$,  and so in the $\sigma (T)$ relation the Horndeski term has
non-trivial effects.  However at large $T$ (corresponding to
large $r_0$, with $T\sim 3 g^2 r_0/(4\pi)$), the $\sigma(T)$ dependence
approaches that obtained in \cite{EMDC}.

\subsection{Dyonic black hole}

We can obtain a more general class of dyonic black hole solutions,
by extending the ansatz for the vector potential in (\ref{ansatz})
to include a magnetic term:
\be
A = a dt + \fft B 2 (x_1 dx_2 - x_2 dx_1) \,.
\ee
We find the dyonic black hole solution is given by
\bea
\phi_1 &=& \lambda x_1 \,, \qquad \phi_2 = \lambda x_2 \,. \cr
a &=&a_0-\fft{q}{r} + \fft{\kappa q (B^2 + q^2)}{30 g^2(4\kappa + \beta \gamma)\,r^5} + \fft{\kappa q \lambda^2}{9(4 \kappa +  \beta \gamma ) g^2 r^3}\,,\cr
\chi' &=& \sqrt{\beta - \fft{\kappa( B^2 +  q^2 + 2 \lambda^2 r^2)}{6 \gamma g^2 r^4}}\,\fft{1}{\sqrt{f}}\,, \cr
f&=& \frac{36 g^4 r^8 (\beta  \gamma +4 \kappa )^2}{\big(\kappa (B^2 + q^2 +2 \lambda ^2 r^2) - 6 g^2 r^4 (\beta  \gamma +4 \kappa )\big)^2} \, h\,,\cr
h&=& g^2 r^2 - \fft{\mu}{r} + \fft{\kappa(B^2 +  q^2)}{ (4\kappa + \beta\gamma)r^2} -
\fft{\kappa^2 (B^2 +  q^2)^2}{60(4\kappa + \beta\gamma)^2g^2 r^6} \cr
&& \qquad  \qquad - \fft{2 \kappa \lambda^2}{4 \kappa +  \beta \gamma} - \fft{\kappa^2 \lambda^4}{3 g^2r^2  (4 \kappa + \beta \gamma )^2} -
\fft{\kappa^2 \lambda^2 (B^2 + q^2) }{9 (4 \kappa + \beta \gamma)^2 g^2 r^4}
\,.
\eea
It is interesting to note that this dyonic solution is rather simply
related to the previous purely electric solution by means of a
replacement in which the quadratic powers of $q$ in (\ref{esl}) are
sent to $q^2+B^2$, while the linear powers of $q$ are left unchanged,
in the sense that one makes the formal replacements
\be
q\rightarrow q\,,\qquad q^2 \rightarrow q^2+B^2\,,\qquad
 q^3\rightarrow q(q^2+B^2)\,.
\ee
The Hawking temperature for the dyonic black hole is given by
\be
T = \frac{ 6 g^2 r_0^4 (\beta  \gamma +4 \kappa ) -   \kappa (B^2 + q^2 + 2 \lambda ^2 r_0^2) }{8 \pi  r_0^3 (\beta  \gamma +4 \kappa )} \,.
\ee

   We are now in a position to calculate the DC conductivity in the
dyonic black hole background. In this case, we turn on
perturbations in both the spatial boundary directions $x^i$,
\bea
\delta g_{tx_1} &=& r^2 \psi_{t1} \,, \quad \delta g_{rx_1} = r^2 \psi_{r1}
\,,\quad \delta g_{tx_2} = r^2 \psi_{t2} \,, \quad
\delta g_{rx_2} = r^2 \psi_{r2} \,, \nn\\
\delta A_{x_1} &=& - E_1 t + a_1 \,, \quad \delta A_{x_2} = - E_2 t + a_2 \,, \quad \delta \phi_1 = \fft{\Phi_1}{\lambda} \,, \quad  \delta \phi_2 = \fft{\Phi_2}{\lambda} \,.
\eea
Following similar methods to those we used in the previous subsection,
we construct a radially-conserved 2-component current
\be
J_i = \kappa \sqrt g F^{rx_i} \,.
\ee
The regularity conditions on the horizon are
\be
a_1' = - \fft{E_1}{\sqrt {hf}} + {\cal O} (1) \,,
\qquad a_2' = - \fft{E_2}{\sqrt {hf}} + {\cal O} (1) \,.
\ee
The currents can be evaluated on the horizon, and we define the
conductivity matrix by
\be
\sigma_{ij} =  \fft{\partial J_i}{\partial E_j} \,,
\quad \text{with} \,, \{ i,j = 1,2 \} \,.
\ee
Explicitly, the conductivity matrix elements are given by
\bea
\sigma_{11} &=& \sigma_{22} = \frac{ \lambda ^2 r_0^2 \left(B^2  +  q^2+\lambda ^2 r_0^2\right)}{B^4 + B^2  \left( q^2+2 \lambda ^2 r_0^2\right)+ \lambda ^4 r_0^4}\,, \cr
\sigma_{12} &=& - \sigma_{21} = \frac{B  q \left(B^2  +  q^2+2 \lambda ^2 r_0^2\right)}{B^4 + B^2  \left( q^2+2 \lambda ^2 r_0^2\right)+\lambda ^4 r_0^4} \,. \label{emdc}
\eea
The Hall angle is defined (for small angles) by
\be
\theta_H = \fft{\sigma_{12}}{\sigma_{11}} =  \frac{B q \left(B^2  +  q^2+2 \lambda ^2 r_0^2\right)}{\lambda ^2 r_0^2 \left(B^2  +  q^2+\lambda ^2 r_0^2\right)}  \,. \label{emhall}
\ee
As in the purely electrically-charged black holes, the inclusion of the
Horndeski
scalar $\chi$ does not modify these transport quantities when they are
expressed in terms of the $r_0$ variable.  In particular the Hall angle 
goes to zero  at high temperature, as $\theta_H\sim 1/T^2$.

\section{Momentum dissipation using Horndeski axions}
\label{EM2H}

\subsection{Dyonic black hole}

   In this section, we consider a system in which the axionic scalars
that provide the momentum dissipation are themselves taken to
have Horndeski couplings rather than minimal couplings to gravity.
The Lagrangian describing the theory is given by
\be
L = \kappa(R-2\Lambda - \ft14 F^2)-
\ft{1}{2}(\alpha g^{\mu\nu}-\gamma G^{\mu\nu})\, \sum_{i=1}^2
 \del_\mu\chi_i\, \del_\nu\chi_i \,.\label{lag3}
\ee
We shall assume that $\alpha$ is positive, and so
for $\gamma=0$ we recover the Einstein-Maxwell theory with two
free axions, proposed in
\cite{EMDC}.  The equations of motion are
\bea
&& \kappa (G_{\mu\nu} +\Lambda g_{\mu\nu} - \fft12 F_{\mu\nu}^2 + \fft18 F^2 g_{\mu\nu})
 -\sum_i^{2}\ft12\alpha \Big(\partial_\mu \chi_i \partial_\nu \chi_i - \ft12 g_{\mu\nu} (\partial\chi_i)^2\Big) \cr
 &&- \sum_i^{2} \ft12\gamma \Big(\ft12\partial_\mu\chi_i \partial_\nu \chi_i R - 2\partial_\rho
\chi_i\, \partial_{(\mu}\chi_i\, R_{\nu)}{}^\rho
- \partial_\rho\chi_i\partial_\sigma\chi_i\, R_{\mu}{}^\rho{}_\nu{}^\sigma  \cr &&
-(\nabla_\mu\nabla^\rho\chi_i)(\nabla_\nu\nabla_\rho\chi_i)+(\nabla_\mu\nabla_\nu\chi_i)
\Box\chi_i + \ft12 G_{\mu\nu} (\partial\chi_i)^2\cr
&&-g_{\mu\nu}\big[-\ft12(\nabla^\rho\nabla^\sigma\chi_i)
(\nabla_\rho\nabla_\sigma\chi_i) + \ft12(\Box\chi_i)^2 -
  \partial_\rho\chi_i\partial_\sigma\chi_i\,R^{\rho\sigma}\big]\Big)  = 0
\,,\nn\\
&&\nabla_\mu \big( (\alpha g^{\mu\nu} - \gamma G^{\mu\nu}) \nabla_\nu\chi_i \big) = 0\,, \qquad
 \nabla_\nu F^{\nu\mu}  =  0\,.
\eea
It is clear that these equations admit a pure AdS vacuum solution where
$R_{\mu\nu}= \Lambda\, g_{\mu\nu}$ and the electromagnetic and scalar fields
vanish.  In this vacuum, the effective kinetic term for the Horndeski axions
$\chi_i$ becomes
\be
L_{(\chi_i, {\rm kin})}= -\ft12(\alpha+\gamma\Lambda) \, \sum_i
(\del\chi_i)^2\,.
\ee
This will be of the standard sign, signifying ghost-freedom, if
$(\alpha +\gamma\Lambda)>0$.  In this paper we shall consider only solutions
for which $\Lambda$ is negative.
 Stability requires that $(\alpha+\gamma\Lambda)$
should be non-negative, but novel features can arise at the critical
point where $(\alpha +\gamma\Lambda)$ vanishes.  (An analogous situation
can also arise in Einstein-Gauss-Bonnet theories, see, e.g., \cite{gwog}.)
Thus $\gamma$ can lies in the range
\be
-\infty < \gamma \le \fft{\alpha}{(-\Lambda)}\,.
\ee

  Typically, the cosmological constant is viewed as a fixed parameter that
is part of the specification of a theory, but it can alternatively arise
as an integration constant for an $n$-form field strength in $n$
dimensions. Thus here we may replace the cosmological constant term in
(\ref{lag3}) by a term
\be
L_{F_\4} = \fft1{4!}\, F_\4^2\,.
\ee
The equation of motion for $F_\4$ can be solved by taking
$F_{\mu\nu\rho\sigma} = \sqrt{-2\Lambda}\, \epsilon_{\mu\nu\rho\sigma}$,
where $\Lambda$ is an arbitrary non-positive constant that acquires an
interpretation as the cosmological constant.  In this new theory, one may
treat the ``cosmological constant'' as a thermodynamic variable, which
has an interpretation as a pressure (see, for example, \cite{karatr,cvgikupo}).
Changing the cosmological constant, i.e. the pressure, can lead to a
phase transition from a stable to an unstable regime as the sign of
$(\alpha+\gamma\Lambda)$ turns negative.  The critical point where
$(\alpha+\gamma\Lambda)$ vanishes gives, as we shall see, some interesting
features in the boundary theory.

We now construct dyonic AdS planar black holes where the two Horndeski
axions are linear
functions of the spatial boundary coordinates $x_i$, i.e.,
\bea
ds^2 &=& - h(r) dt^2 + \fft{dr^2}{f(r)} + r^2 dx^i dx^i \,,\nn\\
A&=& a(r) dt + \fft B 2 (x_1 dx_2 - x_2 dx_1)  \,,\qquad \chi_i =
\lambda x_i \,.
\eea
The equations of motion for the axions are trivially satisfied.
The Maxwell equation implies that
\be
a' = q \sqrt{\fft h f} \, r^{-2} \,,
\ee
where $q$ is an integration constant. With this, the Einstein equations
give
\bea
&& 4 \kappa  r^3 f h'+h \left(2 f \left(\gamma  \lambda ^2+2 \kappa
 r^2\right)+\kappa  (q^2 + B^2) + 4 \kappa  \Lambda  r^4+
  2 \alpha  \lambda ^2 r^2\right) =0 \cr
&& 4 \kappa  r^3 f'+f \left(4 \kappa  r^2-2 \gamma  \lambda ^2\right)+
\kappa  (q^2+B^2)+4 \kappa  \Lambda  r^4+2 \alpha  \lambda ^2 r^2 = 0 \cr
&& h \left(\kappa  r^4 f' h'+f \left(2 \kappa  r^4 h''+
h' \left(2 \kappa  r^3+\gamma  \lambda ^2 r\right)\right)\right) \cr
&& \quad +h^2 \left(f' \left(2 \kappa  r^3+\gamma
\lambda ^2 r\right)-2 \gamma  \lambda ^2 f-\kappa
\left(q^2 + B^2 -4 \Lambda  r^4\right)\right)-\kappa  r^4 f {h'}^2 = 0 \,.
\eea
These equations can be easily solved, leading to the black hole solutions
\bea
h&=& U f \,, \qquad U = e^{\frac{\gamma  \lambda ^2}{2 \kappa  r^2}}\, \cr
a &=& a_0 - \fft{\sqrt{\pi \kappa} q}{\sqrt{\gamma} \lambda}
\text{erfi} (\fft{\sqrt{\gamma}\lambda}{2 \sqrt \kappa r}) \,, \cr
f &=& -\fft{\lambda^2}{6\kappa} (3\alpha + \gamma \lambda)
-\frac{\mu e^{-\frac{\gamma  \lambda ^2}{4 \kappa  r^2}}}{r}-\frac{\Lambda  r^2}{3} \cr
&& + \frac{\sqrt{\pi } e^{-\frac{\gamma  \lambda ^2}{4 \kappa  r^2}}
\text{erfi}\left(\frac{\sqrt{\gamma } \lambda }{2 \sqrt{\kappa } r}\right)
\left(\gamma  \lambda ^4 (3 \alpha +\gamma  \Lambda )+
3 \kappa ^2 (q^2 + B^2)\right)}{12 \sqrt{\gamma } \kappa ^{3/2}
\lambda  r} \,,
\eea
where erfi$(x)$ is the
imaginary error function, defined by erfi$(x)= 2\pi^{-1/2}\,
\int_0^x e^{z^2}\, dz$.
The asymptotic forms of the metric functions near infinity are given by
\bea
-g_{tt} = h(r) \sim - \fft{\Lambda }{3} r^2 - \fft{\lambda^2
(3 \alpha + 2 \gamma \Lambda)}{6 \kappa} - \fft{\mu}{r} +
{\cal O} (\fft{1}{r^2})\,, \nn\\
g^{rr} = f(r) \sim - \fft{\Lambda }{3} r^2 - \fft{\lambda^2
(3 \alpha +  \gamma \Lambda)}{6 \kappa} - \fft{\mu}{r} +
{\cal O} (\fft{1}{r^2})\,,
\eea
which shows that the solution is asymptotic to dS or AdS
for $\Lambda > 0$ or
$\Lambda < 0$ respectively. Since we are interested in the transport
properties of the dual boundary theory, we shall focus on the AdS case,
and so we shall assume  $\Lambda < 0$ in the rest of this section.
The Hawking temperature is given by
\be
T = \frac{
\left(- 4 \kappa  \Lambda  r_0^4 - \kappa(  q^2 + B^2 )  - 2 \alpha
\lambda ^2 r_0^2\right)}{16 \pi  \kappa  r_0^3}\,
\exp\Big(\frac{\gamma  \lambda ^2}{4 \kappa  r_0^2}\Big)\,.\label{TH}
\ee

Although the linearised equations of motion for the Horndeski terms
are of two
derivatives, it is still necessary to check the sign of the kinetic
terms for possible ghost-like behaviour.  The kinetic terms for the
perturbative axions $\delta \chi_i$  are given by
\be
\sum_i^{2} P^{00}\,\delta \dot\chi_i\, \delta\dot \chi_i\,,
\quad \text{with } \, P^{\mu\nu} = -
\ft{1}{2}(\alpha g^{\mu\nu}-\gamma G^{\mu\nu})  \,.
\ee
In order to avoid ghosts, the $P^{00}$ component of $P^{\mu\nu}$,
which is given by
\be
P^{00} = \fft{\alpha - \gamma\, (f r)'}{2 h r^2}=
\frac{\gamma  \kappa  \left(B^2+q^2\right)-2 \gamma ^2 \lambda ^2 f
   +4 \kappa  r^4 (\alpha +\gamma  \Lambda )+
  2 \alpha  \gamma  \lambda ^2 r^2}{8 \kappa  r^4 h} \,,\label{P00}
\ee
should be non-negative, both on and outside the horizon.
The asymptotic form of $P^{00}$ near infinity is given by
\be
P^{00} \sim -\frac{3  (\alpha +\gamma  \Lambda )}{2 \Lambda r^2 } + \frac{\lambda ^2  \left(9 \alpha ^2+12 \alpha  \gamma  \Lambda +5 \gamma ^2 \Lambda ^2\right)}{4 \kappa  \Lambda ^2 r^4} + {\cal O} (\fft{1}{r^5})\,.
\ee
The positivity of $P^{00}$ therefore implies, as a necessary condition,
that $\alpha + \gamma \Lambda \ge 0$ (assuming, as we are, that $\Lambda<0$).
In the case of
equality, $\alpha + \gamma \Lambda = 0$, the leading term of $P^{00}$
vanishes and the asymptotic form of $P^{00}$ becomes simpler, with
\be
P^{00} \sim \frac{\gamma ^2 \lambda ^2 }{2 \kappa r^4} + {\cal O} (\fft{1}{r^6} ) \,,
\ee
which is still greater than zero.
It can then be checked from (\ref{P00})  that $P^{00}$ is indeed
always positive in the region from the horizon to infinity
when $\alpha$ and $\gamma$ are both positive.

\subsection{DC conductivity and Hall angle}

Now, we turn to the calculation of the
DC conductivity of this system. We follow a similar procedure to the
one described in the previous section. Here we shall omit the details of
the calculation, and just present the final results. We begin with the
simpler case where $B=0$, for which we find
the conductivity is given by
\be
\sigma = \kappa +
\frac{4 \kappa ^3 q^2 r_0^2}{\lambda ^2 \left(
4 \kappa  r_0^4 (\alpha +\gamma  \Lambda )+
  2 \alpha  \gamma  \lambda ^2 r_0^2 + \gamma \kappa q^2 \right)} \,.\label{sigcase2}
\ee
When $\gamma=0$, this result reduces to (\ref{cond}).  This demonstrates that the
couplings of the axions for dissipative momenta plays a crucial role in shaping
the conductivity. Although $\sigma$ contains the same ``charge-conjugation symmetric'' term $\kappa$,
as one would expect, it has a very different ``dissipative'' term associated with $\lambda$ that has a richer structure. At large $T$, however, it has the same qualitative behaviour as that of the Einstein-Maxwell case in the high-temperature limit for generic parameters
\be
 \sigma \sim \kappa + \frac{\kappa ^2 q^2}{\lambda ^2 r_0^2
 (\alpha +\gamma  \Lambda )} \sim \kappa +
 \frac{\kappa ^2 q^2 \Lambda^2}{16 \pi^2 \lambda ^2  (\alpha +\gamma  \Lambda )} \fft{1}{T^2} \,.
\ee
On the other hand, at the critical point $\alpha+\gamma\Lambda=0$, the temperature dependence is characteristically different. The denominator of the dissipative term in (\ref{sigcase2})
has three contributions, with the leading-order power of $r_0$ being
proportional to $(\alpha + \gamma \Lambda)$.
When $\alpha + \gamma \Lambda$ is positive, the conductivity rises from
a positive in initial value at
zero temperature, rises to a peak, and then decreases
to a constant value at
high temperature. \footnote{This phenomenon was observed in \cite{mg3}, where massive gravity was used to achieve momentum dissipation.}
Especially, when $\gamma = 0$, the conductivity decreases
monotonically from its initial value as the temperature increases,
behaving much like a normal conductor. If on the other hand
 $\alpha + \gamma \Lambda = 0$, the conductivity
monotonically {\it increases} with temperature,  approaching a constant
in the high-temperature limit.  This behaviour is closer to that of
a semiconductor.
We illustrate the various behaviours in Fig.1, where the parameters are
fixed such that
$\kappa = \alpha = q=1$ and $\lambda = 1/2$. In the left-hand diagram
we fix also $\Lambda=-3$ and display the plots of $\sigma$ versus $T$
for four representative values of $\gamma$. The top curve corresponds to the critical
case $(\alpha+\gamma\Lambda)=0$, while the lower curves correspond
cases with $(\alpha+\gamma\Lambda)>0$.  In the right-hand diagram we
instead fix $\gamma=1/3$ and display plots for various values of $\Lambda$,
again with the critical case $(\alpha+\gamma\Lambda)=0$ being the curve at
the top, with the lower curves having $(\alpha+\gamma\Lambda)>0$.  The
critical case can be thought of as representing a phase transition where
the high-temperature behaviour of the material changes from that of a
metal ($\sigma$ falls to a small constant $\kappa$ as $T$ increases)
to a semiconductor ($\sigma$
rises to a limiting value as $T$ increases) in the critical case.  From the
bulk point of view, the transition can be viewed as being induced when the
pressure ($\sim (-\Lambda)$) becomes sufficiently large.

   It is interesting to note that in the left-hand diagram in Fig.~1, all
the conductivity curves originate from the same value when $T=0$.  The
reason for this can be seen from the expressions for the temperature and
the conductivity, namely
\bea
T &=& -\frac{e^{\frac{\gamma  \lambda ^2}{4 \kappa  r_0^2}}
\, \left(4 \kappa  \Lambda  r_0^4+
     2 \alpha  \lambda ^2 r_0^2 +
      \kappa q^2\right)}{16 \pi  \kappa  r_0^3} \,,\label{Teq}\\
\sigma &=&\kappa +\frac{4 \kappa ^3 q^2 r_0^2}{\gamma
\left(4 \kappa  \Lambda  r_0^4
    +2 \alpha  \lambda ^2 r_0^2 + \kappa q^2\right)  +4 \alpha  \kappa  r_0^4}
\,.\label{sigeq}
\eea
The temperature becomes zero when the factor in parentheses in (\ref{Teq})
vanishes, and then (\ref{sigeq}) implies that the corresponding
zero-temperature conductivity is given by
\be
\sigma(0)  = \kappa +\frac{\kappa ^2 q^2}{\alpha  \lambda ^2 r_0^2} \,,
\ee
with $r_0$ being given by
\be
r_0^2 = \fft{\alpha \lambda^2 +\sqrt{\alpha^2\lambda^4
    - 4 \kappa^2 q^2 \Lambda}}{4\kappa (-\Lambda)}\,.
\ee
Thus at fixed $\Lambda$, with $\kappa$, $\alpha$, $q$ and $\lambda$
also fixed as in left-hand diagram,
the zero-temperature conductivity is independent of $\gamma$. By contrast,
if $\gamma$ is fixed instead of $\Lambda$, as in the right-hand
diagram, the zero-temperature conductivity does depend on $\Lambda$.

We have not included plots for values of the parameters for which
$\alpha + \gamma \Lambda$ is negative. Here, the dissipative part of the
conductivity
can be negative, and for a range of temperatures the full expression for the
conductivity can be negative or divergent.  This suggests
an unphysical instability, and is in fact consistent with our
previous observation that the Horndeski axions acquire ghost-like kinetic
terms when $\alpha + \gamma \Lambda$ is negative.

\begin{figure}[htp]
\begin{center}
\includegraphics[width=200pt]{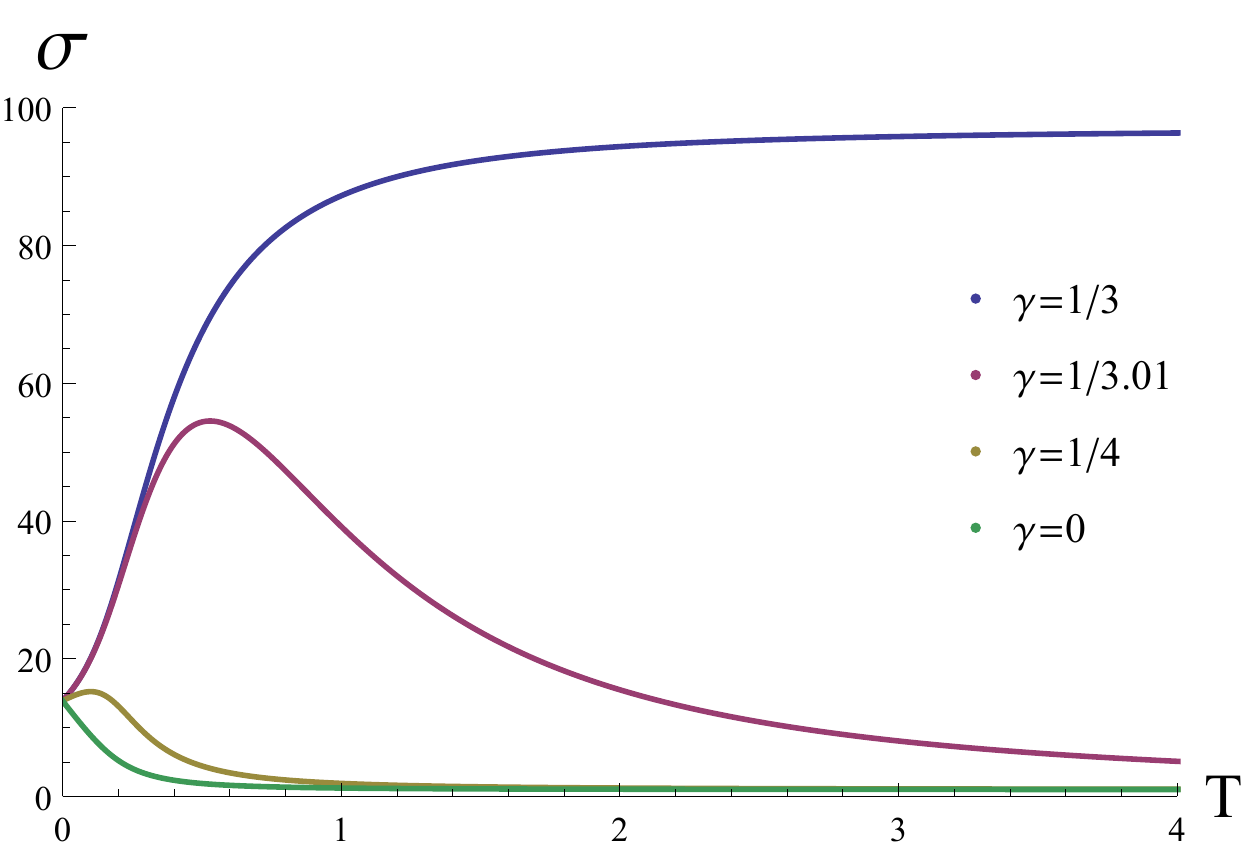}
\includegraphics[width=200pt]{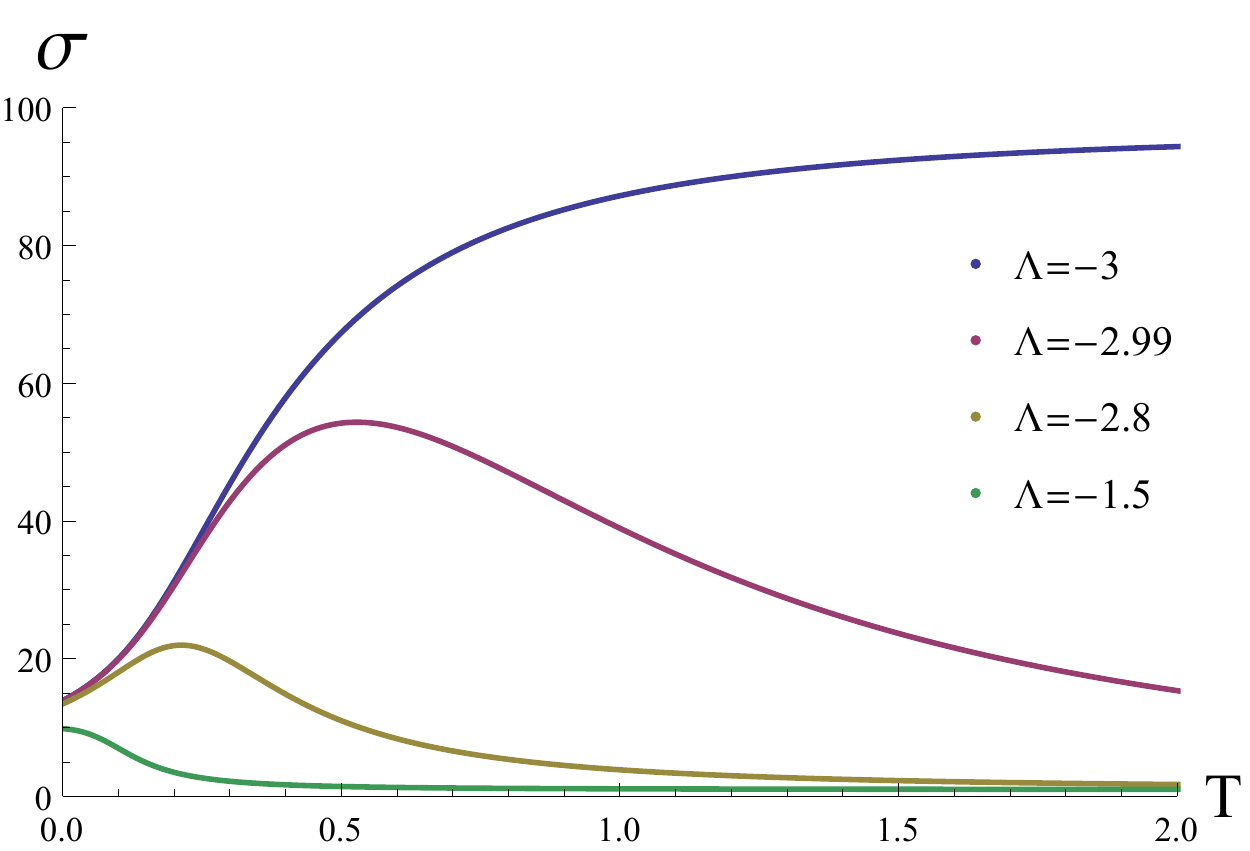}
\end{center}
\caption{\setstretch{1.0}\small\it Plots of the conductivity $\sigma$ versus
temperature, for various parameter choices.  In each diagram we have set
$\kappa =\alpha = q=1$ and $\lambda = 1/2$.  In the left-hand
diagram we fix $\Lambda=-3$ and take various choices for the parameter
$\gamma$.  The top line has the
critical value $\gamma = 1/3$, for which $(\alpha+\gamma\Lambda)=0$.
In the right-hand diagram we instead fix $\gamma=1/3$ and take
various choices for the parameter
$\Lambda$.  Again, the top line corresponds to the
critical value.  In both diagrams, the lower lines all correspond to
$(\alpha + \gamma\Lambda)>0$, and they 
approach $\kappa$ (which we have set equal to 1 for the purposes 
of these plots) at large $T$.} \label{figure1}
\end{figure}

The case when  $B \ne 0$ is considerably more complicated,
and we shall not present the general expression for the conductivity
matrix here. However, in the high temperature limit we find that it takes
the form for $\alpha + \gamma \Lambda>0$
\bea
\sigma_{11} &=& \sigma_{22} \sim ~ \kappa + \frac{\kappa ^2 \Lambda^2
\left(q^2 - B^2\right)}{16 \pi^2 \lambda^2  (\alpha +\gamma  \Lambda )T^2} \,,\nn\\
\sigma_{12} &=& - \sigma_{21} \sim   \frac{ \kappa ^2 \Lambda ^2 q B}{8 \pi ^2 \lambda ^2 (\alpha +\gamma  \Lambda )T^2} \cr
 && \qquad \quad + \frac{ \kappa  \Lambda ^3 q B \left(2 \alpha ^2 \lambda ^4+2 \alpha  \gamma  \lambda ^4 \Lambda +\Lambda  \left(-3 B^2 \kappa ^2+\gamma ^2 \lambda ^4 \Lambda +\kappa ^2 q^2\right)\right)}{256 \pi ^4 \lambda ^4 (\alpha +\gamma  \Lambda )^2 T^4}  \,,
\eea
and the Hall angle is given by
\be
\theta_H \sim \frac{ \kappa  \Lambda ^2 q B}{8 \pi ^2 \lambda ^2
(\alpha +\gamma  \Lambda )T^2} +
\frac{ \Lambda ^3 q B \left(2 \alpha ^2 \lambda ^4+
2 \alpha  \gamma  \lambda ^4 \Lambda +\Lambda
\left(-B^2 \kappa ^2+\gamma ^2 \lambda ^4
\Lambda -\kappa ^2 q^2\right)\right)}{256 \pi ^4
\lambda ^4 (\alpha +\gamma  \Lambda )^2 T^4} \,.
\ee
At the critical point $\alpha + \gamma \Lambda = 0$, the conductivity
and Hall angle at high temperature become
\bea
\sigma_{11}  = \sigma_{22} && \sim \quad  \frac{\gamma ^2 \kappa  \lambda ^4 \Lambda  \left(\gamma ^2 \lambda ^4 \Lambda -2 \kappa ^2( q^2 + B^2 ) \right)}{4 B^4 \kappa ^4+4 B^2 \left(\kappa ^4 q^2-\gamma ^2 \kappa ^2 \lambda ^4 \Lambda \right)+\gamma ^4 \lambda ^8 \Lambda ^2} + \,{\cal O} (\fft{1}{T^2}) \,, \cr
\sigma_{12} = - \sigma_{21} &&  \sim \quad \frac{4 B \kappa ^3 q \left( \kappa ^2 ( q^2 + B^2 )-\gamma ^2 \lambda ^4 \Lambda \right)}{4 B^4 \kappa ^4+4 B^2 \left(\kappa ^4 q^2-\gamma ^2 \kappa ^2 \lambda ^4 \Lambda \right)+\gamma ^4 \lambda ^8 \Lambda ^2} + \,{\cal O} (\fft{1}{T^2}) \,,\cr
\theta_H && \sim \quad \frac{4 B \kappa ^2 q \left(\kappa ^2 (B^2 + q^2)-\gamma ^2 \lambda ^4 \Lambda \right)}{\gamma ^2 \lambda ^4 \Lambda  \left(-2 B^2 \kappa ^2+\gamma ^2 \lambda ^4 \Lambda -2 \kappa ^2 q^2\right)} + \,{\cal O} (\fft{1}{T^2}) \,.
\eea
In particular, the Hall angle approaches a constant at large $T$.
At $T=0$, on the other hand, which occurs when the factor in
parentheses in the numerator in (\ref{TH}) vanishes,
the conductivity and Hall angle become
\bea
\sigma_{11}  = \sigma_{22} &=& \frac{\alpha  \kappa  \lambda ^2 r_0^2 \left(B^2 \kappa +\kappa  q^2+\alpha  \lambda ^2 r_0^2\right)}{B^4 \kappa ^2+B^2 \kappa  \left(\kappa  q^2+2 \alpha  \lambda ^2 r_0^2\right)+\alpha ^2 \lambda ^4 r_0^4}  \,, \cr
\sigma_{12} = - \sigma_{21} &=& \frac{B \kappa ^2 q \left(B^2 \kappa +\kappa  q^2+2 \alpha  \lambda ^2 r_0^2\right)}{B^4 \kappa ^2+B^2 \kappa  \left(\kappa  q^2+2 \alpha  \lambda ^2 r_0^2\right)+\alpha ^2 \lambda ^4 r_0^4}  \,, \cr
\theta_H (0) &=& \frac{ \kappa  q B \left(B^2 \kappa +\kappa  q^2+2 \alpha  \lambda ^2 r_0^2\right)}{\alpha  \lambda ^2 r_0^2 \left(B^2 \kappa +\kappa  q^2+\alpha  \lambda ^2 r_0^2\right)} \,,
\eea
where
\be
r_0 = \frac{\alpha  \lambda ^2+\sqrt{\alpha ^2 \lambda ^4-4 \kappa ^2 \Lambda  \left(B^2+q^2\right)}}{ - 4 \kappa  \Lambda } \,.
\ee
It is of interesting to note that at $T=0$, both $\sigma$'s and $\theta_H$ are independent of $\gamma$. This implies that in the zero temperature limit the DC conductivities and Hall angle are the same as those in the Einstein-Maxwell case (\ref{emdc},\ref{emhall}), if the results are expressed in terms of the horizon radius $r_0$ and we set $\alpha \rightarrow\kappa$.

\section{Conclusions}

    In this paper, we studied two four-dimensional
gravity theories involving scalar fields
with non-minimal Horndeski-type couplings to gravity. We first considered
Einstein-Maxwell gravity with one non-minimally coupled Horndeski
axion and two minimally coupled axions. The two minimally coupled axions
have linear dependence on the spatial boundary coordinates, and
they generate momentum dissipation in the standard way. We constructed
a charged AdS planar black hole in the theory, and calculated the holographic
DC conductivity in the dual field theory. Interestingly, although
the Horndeski
scalar in these solutions plays a role in determining the geometry of
the black hole background, it does not contribute directly
to the conductivity.  To be precise, if
written in terms of the horizon radius $r_0$ the conductivity is the same
as that in Maxwell-Einstein gravity.

In the second model, we used two Horndeski axions, non-minimally coupled to
Einstein-Maxwell gravity, to drive the momentum dissipation. We obtained
a static AdS black hole solution in the theory. We analyzed the kinetic terms
of the axion perturbations, and showed that the theory has a
critical point at $\alpha + \gamma \Lambda=0$. When
$\alpha+ \gamma \Lambda < 0$, the kinetic
terms of the axion perturbations become negative, implying that
the excitations become ghost-like.  We then obtained the
conductivity in the dual boundary theory, and found that the conductivity
has two terms, a ``charge-conjugation symmetric'' term and ``dissipative''
term as usual. However, the dissipative term has richer features than in a
standard minimally-coupled theory. At the critical point
$\alpha + \gamma \Lambda = 0$, the conductivity increases monotonically
as a function of temperature, which is typical of the behaviour in
a semiconductor. When  $\alpha + \gamma \Lambda > 0$, on the other hand,
the conductivity rises to a maximum then falls, finally approaching
a constant. In the special case $\gamma =0$, corresponding to turning off the
Horndeski modification of the usual minimal coupling of the axions,
the conductivity decreases monotonically with temperature, and the behavior
is more like a normal conductor. We chose a set of parameters in the paper
and plotted the conductivity versus temperature curves for various values of
$\gamma$, in Fig.~1. We showed that from the critical point
$\gamma = - \alpha/\Lambda$ to the special case $\gamma = 0$,
the behavior of the conductivity as a function of temperature changes
from that reminiscent of a semiconductor to that of a normal conductor.

Momentum dissipation is the key for obtaining finite holographic DC
conductivity.
While free axions provide one of the simplest models for such a mechanism,
the resulting DC conductivity generally tends to have a fairly
simple structure whose qualitative features are independent of the parameters.  Our work demonstrated that using non-minimally
coupled axions in the momentum-dissipation mechanism can lead to a
much richer pattern of holographic DC conductivities.

\section*{Acknowledgements}

We are grateful to Sera Cremonini for discussions.
H-S.L.~is supported in part by NSFC grants No.~11305140, 11375153,
11475148, 11675144 and CSC scholarship No.~201408330017.  The work of
H.L.~is supported in part by NSFC grants No.~11475024, No.~11175269
and No.~11235003. C.N.P.~is supported in part by DOE grant DE-FG02-13ER42020.

\appendix

\section{Higher dimensional case}
In this section, we generalise the theory in section 2 to include $N$
$p$-form fields in arbitrary dimension,
\be
L = \kappa\Big(R-2\Lambda - \fft14 F^2 -\sum_{i=1}^N \fft{1}{2 p!} \big({\cal F}^i_{\sst{(p)}} \big)^2\Big)-
\fft{1}{2}(\alpha g^{\mu\nu}-\gamma G^{\mu\nu})
\, \del_\mu\chi\, \del_\nu\chi \,,
\ee
where $\kappa$, $\alpha$ and $\gamma$ are coupling constants,
$G_{\mu\nu} \equiv R_{\mu\nu} - \fft12 R g_{\mu\nu}$ is the Einstein tensor,
$F=dA$ is the electromagnetic field strength and ${\cal F}^i = d {\cal A}^i$ is one of the form fields which span all spacial dimension with $N p = n-2 $. The equations of motion are given by
\bea
&&\kappa (G_{\mu\nu} +\Lambda g_{\mu\nu} - \fft12 F_{\mu\nu}^2 + \fft 18 F^2 g_{\mu\nu})  + \sum_{i = 1} ^N \big[ - \fft{\kappa}{2 (p-1)!} \big( {\cal F}^i \big)^2_{\mu\nu} + \fft{\kappa}{4 p!} \big( {\cal F}^i  \big)^2 g_{\mu\nu}\big]\cr
&& -\ft12\alpha \Big(\partial_\mu \chi \partial_\nu \chi - \ft12 g_{\mu\nu} (\partial\chi)^2\Big)-\ft12\gamma \Big(\ft12\partial_\mu\chi \partial_\nu \chi R - 2\partial_\rho
\chi\, \partial_{(\mu}\chi\, R_{\nu)}{}^\rho \cr
&&- \partial_\rho\chi\partial_\sigma\chi\, R_{\mu}{}^\rho{}_\nu{}^\sigma -
(\nabla_\mu\nabla^\rho\chi)(\nabla_\nu\nabla_\rho\chi)+(\nabla_\mu\nabla_\nu\chi)
\Box\chi + \ft12 G_{\mu\nu} (\partial\chi)^2\cr
&&-g_{\mu\nu}\big[-\ft12(\nabla^\rho\nabla^\sigma\chi)
(\nabla_\rho\nabla_\sigma\chi) + \ft12(\Box\chi)^2 -
  \partial_\rho\chi\partial_\sigma\chi\,R^{\rho\sigma}\big]\Big)  = 0\,,\nn\\
&& \nabla_\mu \big( (\alpha g^{\mu\nu} - \gamma G^{\mu\nu}) \nabla_\nu\chi\big) = 0\,, \qquad
\nabla_\nu F^{\nu\mu}  =  0\,,  \quad \nabla_\nu {\cal F}_i^{\nu\mu_1\cdots \mu_{p-1}}  =  0
\eea

We consider static planar black hole ansatz
\bea
&&ds^2 = - h(r) dr^2 + \fft{dr^2}{f(r)} + r^2 dx^i dx^i \,, \cr
&&\chi = \chi(r) \,, \quad A=a(r) \,dt\,, \quad {\cal F}^i = \lambda dx^i_1 \wedge \dots \wedge dx^i_p \,.
\eea
where $\lambda$ is a constant.
The Maxwell's equation can be used to express the electrical potential in terms of metric functions
\be
a' = q \sqrt{\fft h f} r^{2-n} \,,
\ee
where $q$ is an integration constant.
And the equation of motion for scalar can be written as
\be
\Big(r^{n-4} \sqrt{\fft{f}{h}}
\Big( \gamma \big( (n-2) r f h' + (n-2)(n-3) f h\big)
    -2\alpha r^2 h \Big)\chi'\Big)'=0\,.
\ee
We focus on a special class of solution, as what we did in section 2, by letting
\be
\gamma \big( (n-2) r f h' + (n-2)(n-3) f h\big)
   -2\alpha r^2 h=0\,.
\ee
Under these setup, we can obtain the black hole solution
\bea
a &=& a_0 -\frac{q  }{(n-3) r^{n-3} }+\frac{\kappa  q^3  }{g^2 (3 n-7) (n-2) (n-1) (\beta  \gamma +4 \kappa ) r^{3 n - 7} } \cr
&&+\frac{N \kappa \lambda ^2 q }{g^2 (n-2) (n-1) (n+2 p-3) (\beta  \gamma +4 \kappa )  r^{n+2 p-3} } \,, \cr
\chi' &=&  \sqrt{ \beta -\frac{\kappa ( q^2 + N \lambda ^2 r^{2 n -2 p - 4})}{\gamma  g^2 \left(n^2-3 n+2\right) r^{2 n - 4}  }  } \,\fft{1}{\sqrt{f}}\,, \cr
f &=& \frac{g^4 (n-2)^2 (n-1)^2 (\beta  \gamma +4 \kappa )^2 r^{4n - 8}} { \left(  \kappa  q^2 -g^2 \left(n^2-3 n+2\right) (\beta  \gamma +4 \kappa ) r^{2 n -4} + N \kappa \lambda ^2 r^{2 n - 2 p -4} \right)^2} \, h\,,\cr
h &=& g^2 r^2 -\fft{\mu}{  r^{n-3} } +\frac{2 \kappa  q^2  }{(n-3) (n-2) (\beta  \gamma +4 \kappa ) r^{2 n - 6}}\nn\\
&&+\frac{\kappa ^2 q^4  }{g^2 (7-3 n) (n-2)^2 (n-1) (\beta  \gamma +4 \kappa )^2 r^{4 n  - 10}} -\frac{2 N \kappa \lambda ^2 }{(n-2) (n-2 p-1) (\beta  \gamma +4 \kappa ) r^{2 p - 2} }\nn\\
 && + \frac{N^2 \kappa^2 \lambda ^4  }{g^2 (n-2)^2 (n-1) (n-4 p-1) (\beta  \gamma +4 \kappa )^2 r^{4 p - 2}} \cr
&& -\frac{2 N \kappa^2  \lambda ^2 q^2  }{g^2 (n-2)^2 (n-1) (n+2 p-3) (\beta  \gamma +4 \kappa )^2 r^{2 (n+p) - 6 }  } \,,
\eea
with parameters under constraint
\be
\alpha=\ft12 (n-1)(n-2) g^2 \gamma\,,\qquad
\Lambda = -\ft12 (n-1)(n-2)g^2 \Big(1 + \fft{\beta\gamma}{2\kappa}\Big)\,.
\ee

\section{Einstein-Maxwell-Dilaton theory with Horndeski axions}

In section \ref{EM2H}, we studied the theory of Einstein-Maxwell gravity
with two non-minimally coupled Horndeski axions.  Here, we give
a generalisation in which we include also a dilatonic scalar field with
an exponential coupling to the Maxwell field, and exponential potential
terms.  The Lagrangian is given by
\be
L = \kappa[R-2\Lambda e^{\delta_0 \phi} -
 2 V_0 e^{\delta_2 \phi} - \ft12 (\partial \phi)^2 -
\ft14 e^{\delta_1 \phi} F^2]
 - \sum_i^{2}
\ft{1}{2}(\alpha g^{\mu\nu}-\gamma G^{\mu\nu})
\, \del_\mu\chi_i\, \del_\nu\chi_i \,.
\ee
where $\delta_0 \,, \delta_1 \,, \delta_2 \,, V_0 \,, \kappa \,, \gamma \,,$
and $\alpha$ are constants.  The second potential term, with coefficient
$V_0$, is required for the case where a magnetic field is included.
The equations of motion are given by
\bea
&& \kappa (G_{\mu\nu} +(\Lambda e^{\delta_0 \phi} + V_0 e^{\delta_2 \phi} ) g_{\mu\nu} - \ft12 \partial_\mu \phi \partial_\nu \phi + \ft14 (\partial \phi)^2 g_{\mu\nu} - \ft12 e^{\delta_1 \phi} F_{\mu\nu}^2 + \ft18 e^{\delta_1 \phi} F^2 g_{\mu\nu}) \cr
 && - \sum_i^{2}\ft12\alpha \Big(\partial_\mu \chi_i \partial_\nu \chi_i - \ft12 g_{\mu\nu} (\partial\chi_i)^2\Big) \cr
 &&- \sum_i^{2} \ft12\gamma \Big(\ft12\partial_\mu\chi_i \partial_\nu \chi_i R - 2\partial_\rho
\chi_i\, \partial_{(\mu}\chi_i\, R_{\nu)}{}^\rho
- \partial_\rho\chi_i\partial_\sigma\chi_i\, R_{\mu}{}^\rho{}_\nu{}^\sigma  \cr &&
-(\nabla_\mu\nabla^\rho\chi_i)(\nabla_\nu\nabla_\rho\chi_i)+(\nabla_\mu\nabla_\nu\chi_i)
\Box\chi_i + \ft12 G_{\mu\nu} (\partial\chi_i)^2\cr
&&-g_{\mu\nu}\big[-\ft12(\nabla^\rho\nabla^\sigma\chi_i)
(\nabla_\rho\nabla_\sigma\chi_i) + \ft12(\Box\chi_i)^2 -
  \partial_\rho\chi_i\partial_\sigma\chi_i\,R^{\rho\sigma}\big]\Big)  = 0\,,\nn\\
&&\nabla_\mu \big( (\alpha g^{\mu\nu} - \gamma G^{\mu\nu}) \nabla_\nu\chi_i \big) = 0\,, \qquad
E_A^\mu\equiv \nabla_\nu (e^{\delta_1 \phi} F^{\nu\mu} )  =  0\,,  \nn\\
&& \Box \phi - 2 \Lambda \delta_0 e^{\delta_0 \phi}  - 2 V_0 \delta_2 e^{\delta_2 \phi}- \ft14{\delta_1} e^{\delta_1 \phi} F^2  = 0 \,.
\eea
We consider the static planar black hole in four dimensions
\bea
ds^2 &=& - h(r) dr^2 + \fft{dr^2}{f(r)} + r^2 dx^i dx^i \,, \cr
\chi_i &=& \lambda x_i \,, \quad A=a(r) dt + \fft B 2 ( x_1 dx_2 - x_2 dx_1)
\,, \quad  \phi = \beta \log r \,,
\eea
where $\lambda \,, \beta $ and $B$  are constants.
The Maxwell equation implies
\be
a' = q \sqrt{\fft h f} r^{-2 - \delta_1 \beta} \,.
\ee
We find that there are two inequivalent classes of solutions,
where the parameters $(\delta_0,\delta_1,\delta_2)$  are  given by
\bea
\hbox{Class 1}:&& \delta_0=-\fft{2}{\beta}\,,\qquad
\delta_1= \fft{\beta}{2}\,,\qquad \delta_2= \fft{\beta}{2}-\fft{4}{\beta}\,,
\nn\\
\hbox{Class 2}:&& \delta_0 =-\fft{2}{\beta}\,,\qquad
 \delta_1= -\fft{2}{\beta}\,,\qquad \delta_2= -\fft{6}{\beta}\,.
\eea
In both classes we have
\be
h= U f \,,
\qquad U = r^{\ft12\beta^2}\,
e^{\frac{\gamma  \lambda ^2}{2 \kappa  r^2}} \nn\,.
\ee
For class 1 we find
\bea
f&=& e^{-\fft{\gamma \lambda^2}{4\kappa r^2}}\,\Big[
\fft{\alpha \lambda^2}{\kappa \,(\beta^2-4)}\,
\Ei\left(\ft32 +\ft18\beta^2,-\fft{\gamma\lambda^2}{4\kappa r^2}\right) +
\fft{B^2}{2(\beta^2-4)}\, r^{-2+\ft12 \beta^2}\,
\Ei\left(\ft12 +\ft38\beta^2,-\fft{\gamma\lambda^2}{4\kappa r^2}\right)
\nn\\
&&\qquad\qquad -\ft18 q^2 \, r^{-2-\ft12\beta^2}\,
\Ei\left(\ft12 -\ft18\beta^2,-\fft{\gamma\lambda^2}{4\kappa r^2}\right)
 -\mu\, r^{-1-\ft14 \beta^2} \Big]\,,
\eea
with parameters
\be
\Lambda = \frac{\alpha  \beta ^2 \lambda ^2}{2 \kappa(4 -\beta ^2) }
\,, \qquad V_0 = \fft{\beta^2 B^2}{4 (4-\beta^2)} \,.
\ee
$\Ei$ is the exponential integral, defined by
\be
\Ei(z,x)= \int_{1}^\infty t^{-z}\,
e^{-x t}\, dt
= \Gamma(1-z)\, x^{z-1}-
\sum_{n\ge 0} \fft{(-x)^n}{n!\, (n+1-z)}\,.\label{Eidef}
\ee
The Hawking temperature for the class 1 solutions is given by
\be
T = \frac{e^{\frac{\gamma  \lambda ^2}{4 \kappa  r_0^2}}\, r_0^{\ft14\beta^2}\,
\left(4 B^2 \kappa  r_0^{\beta^2}-\left(\beta^2-4\right) \kappa
q^2+8 \alpha  \lambda^2 r_0^{\frac{\beta^2}{2}+2}\right)}{16 \pi
\left(\beta^2-4\right) \kappa  r_0^3}\,.
\ee
The positivity of temperature require $\beta^2 > 4 $. In the large $r_0$ limit, the temperature approaches
\be
T \sim \frac{B^2 r_0^{(\beta ^2-3) } }{4 \pi  \left(\beta ^2-4\right)} \,.
\ee
For the class 2 solutions we find
\bea
f&=& e^{-\fft{\gamma \lambda^2}{4\kappa r^2}}\,\Big[
\fft{(\alpha \lambda^2+ \kappa q^2)}{\kappa \,(\beta^2-4)}\,
\Ei\left(\ft32 +\ft18\beta^2,-\fft{\gamma\lambda^2}{4\kappa r^2}\right) +
\fft{B^2}{(\beta^2-12)\, r^4}\,
\Ei\left(-\ft12 +\ft18\beta^2,-\fft{\gamma\lambda^2}{4\kappa r^2}\right)
\nn\\
&&\qquad \qquad  -\mu\, r^{-1-\ft14 \beta^2} \Big]\,,
\eea
with parameters
\be
\Lambda= -\fft{2\alpha \beta^2\, \lambda^2 + (\beta^2+4)\, \kappa q^2}{
   4\kappa (\beta^2-4)}\,,\qquad
V_0= \fft{B^2\, (\beta^2-4)}{4(12-\beta^2)}\,.
\ee
The Hawking temperature for the class 2 solutions is given by
\be
T= e^{\frac{\gamma  \lambda ^2}{4 \kappa  r_0^2}}\, r_0^{\ft14\beta^2}\,
 \fft{(12-\beta^2) (\alpha \lambda^2 + \kappa q^2) r_0^4 -
   B^2\, \kappa\, (\beta^2-4)}{2(\beta^2-4)(12-\beta^2)\, \kappa \pi r_0^5}\,.
\ee
It can be seen from the series expansion for the exponential integral
function given in (\ref{Eidef}) that in the case of the class 2 solutions,
the non-integer powers of $r^{-1}$ that arise, for generic values of
$\beta$, in the large-$r$ expansion of the metric function $f$ can be removed
altogether if the constant $\mu$ is chosen to be given by
\be
\mu= -\Gamma(\ft12-\ft18 \beta^2)\,
\Big(-\fft{\gamma\lambda^2}{4\kappa}\Big)^{\ft12 +\ft18\beta^2}\, \Big[
 \fft{2 B^2 \kappa^2 (\beta^2-4)}{\gamma^2\, \lambda^2\, (\beta^2-12)}
   + \fft{8(\alpha\lambda^2 + \kappa q^2)}{(\beta^4-16)\kappa}\Big]\,.
\ee

\end{document}